\documentclass{vgtc}                          

\graphicspath{{figures/}{pictures/}{images/}{./}} 
\usepackage{enumitem}
\usepackage{times} 
\usepackage{float}

\usepackage{tabu}                      
\usepackage{booktabs}                  
\usepackage{lipsum}                    
\usepackage{mwe}                       

\usepackage{mathptmx}                  
\usepackage{balance}

\onlineid{3853}

\vgtccategory{Research}

\vgtcinsertpkg

\title{Preliminary Evaluation of Transition-Aware Controller Locomotion Adaptations for Supporting Postural Stability in VR}

\author{Ramisa Fariha Joyee\thanks{e-mail: ramisafariha.joyee123@gmail.com}\\ %
        \scriptsize Kennesaw State University %
\and M. Rasel Mahmud\thanks{e-mail: m.raselmahmud1@gmail.com (Corresponding Author)}\\ %
     \scriptsize Kennesaw State University %
}

\teaser{
  \centering
  \includegraphics[width=\linewidth, height=0.26\textheight]{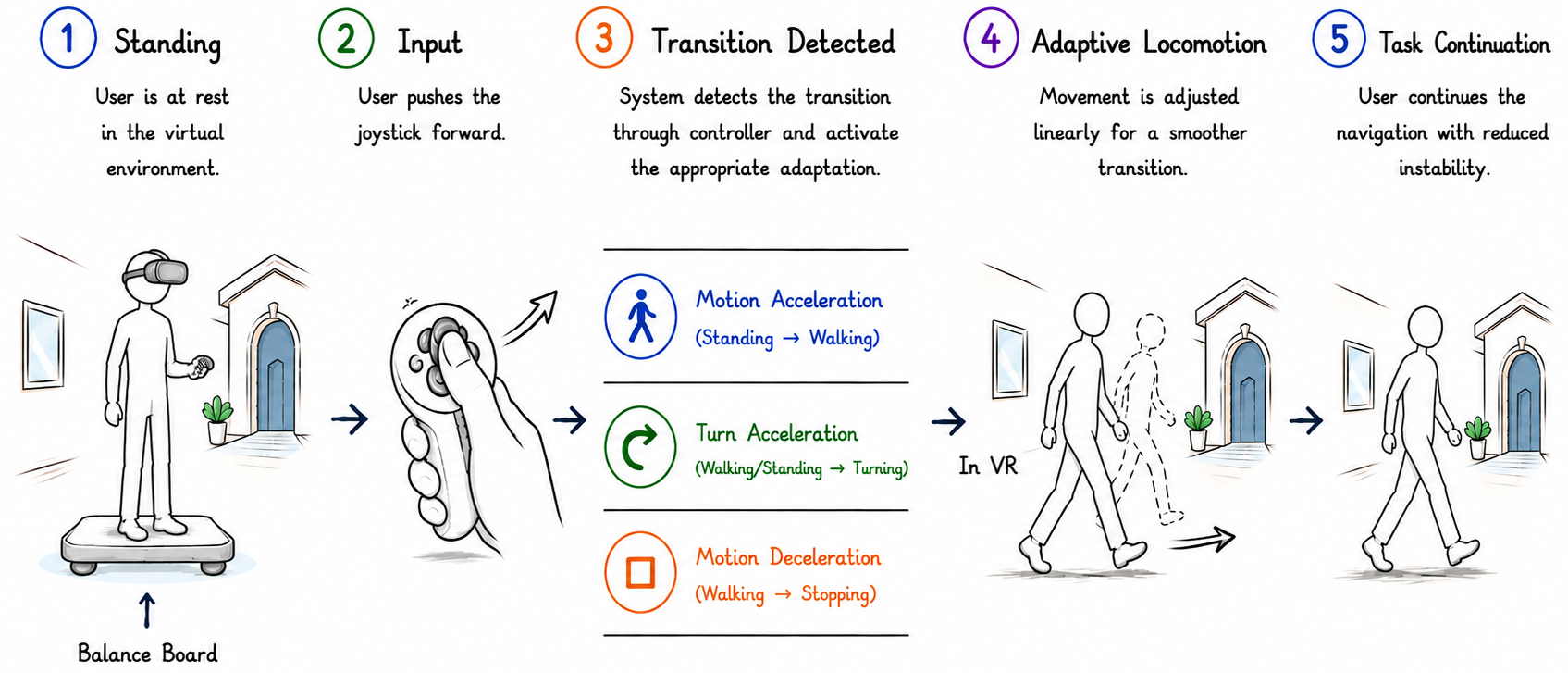}
  \includegraphics[width=\linewidth]{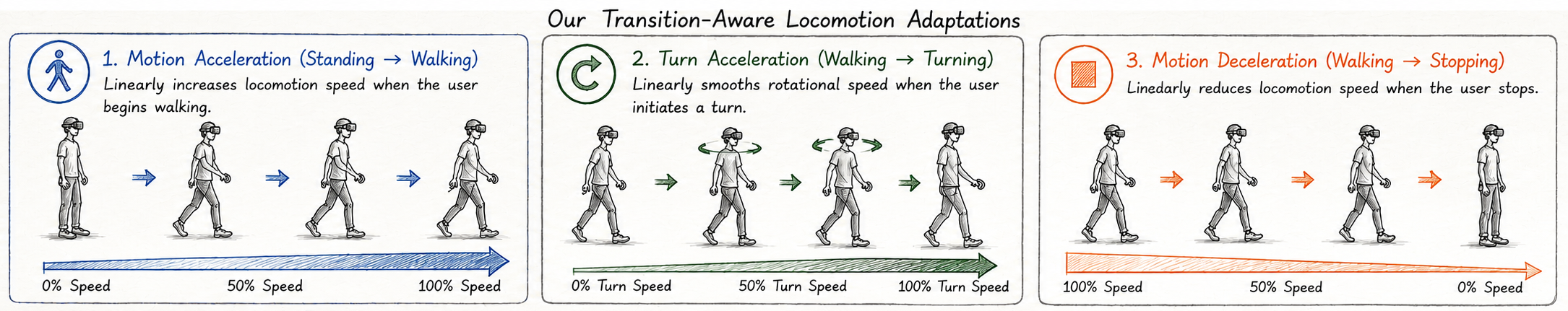}
  \caption{A conceptual depiction showing proposed transition-aware locomotion adaptations (motion acceleration, turn acceleration, motion deceleration) during different virtual body-state transitions (e.g., standing-to-walking, turning, walking-to-stopping) in VR.}
  \label{fig:flow_picture_adaptation}
}

\abstract{
    We present three locomotion adaptation approaches: Motion Acceleration, Turn Acceleration, and Motion Deceleration to improve postural stability during body-state transitions in virtual reality (VR). The system detects standing-to-walking, turning, and walking-to-stopping transitions and applies adaptive locomotion smoothing. Motion Acceleration gradually increases locomotion speed when users begin walking, Turn Acceleration smooths rotation while turning, and Motion Deceleration gradually reduces movement speed before stopping. We evaluated these techniques in a virtual navigation task using objective and subjective balance measures. Preliminary results show reduced center of pressure (COP) velocity and improved balance confidence. These findings suggest that locomotion adaptations can improve balance and navigation experience.
} 


\begin{document}



\maketitle

\section{Introduction} 
    Human locomotion is inherently continuous. People do not suddenly start walking or turning; rather, they gradually accelerate when they begin walking, smoothly adjust their movement while turning, and naturally decelerate before coming to a stop. In contrast, hand controller-based locomotion in virtual reality (VR) often produces abrupt transitions between these movement states, creating sudden changes in visual motion that do not reflect natural human movement. As users repeatedly start, turn, and stop during navigation, these abrupt transitions may require additional postural adjustments to maintain balance. Maintaining stable posture is particularly important during navigation or complex spatial tasks, where users continuously process spatial information, make directional decisions, and coordinate body movements within the virtual environment \cite{horejsi_2025_vrlocomotion, keshner_2004_posturalresponse}. Research on postural control explains that maintaining human balance depends on the integration of visual, vestibular, and proprioceptive sensory inputs \cite{missen_2024_sensorysystem}. According to the ecological theory of motion sickness \cite{riccio_1991_ecological, stoffregen_1998_postural}, prolonged postural instability contributes to the development of motion sickness. Therefore, reducing postural instability during locomotion transitions may improve user comfort in VR. Since VR environments heavily rely on visually generated motion cues, inconsistencies between perceived virtual motion and physical body movement can disrupt normal sensory integration and negatively affect postural stability and spatial orientation \cite{virk_2006_improve_posture}. Most existing approaches primarily optimize continuous locomotion, such as movement speed, rendering quality, navigation guidance, or user interaction \cite{chi_eeg, shi_2023_foveated, wang2021fuzzy}. Comparatively little attention has been given to the transition periods between body states, including standing-to-walking, walking-to-turning, and walking-to-stopping. These transitions occur frequently during navigation and represent moments when users experience abrupt changes in optic flow and movement dynamics. Yet current VR locomotion systems generally treat them as instantaneous events without adapting locomotion behaviour.

    To address this gap, we propose three transition-aware locomotion adaptations that dynamically smooth different virtual body-state transitions during VR navigation. Figure \ref{fig:flow_picture_adaptation} illustrates the overall workflow of the proposed transition-aware locomotion system, including transition detection and the three adaptation strategies. Motion Acceleration gradually increases locomotion speed when users initiate movement, Turn Acceleration progressively increases rotational speed at the onset of turning, and Motion Deceleration gradually reduces movement speed before stopping. Rather than modifying locomotion continuously, the proposed adaptations are activated only during specific transition events. This proposed approach allows users to retain conventional locomotion while reducing abrupt changes in visual motion.
    The main contributions of this work are the following:
    \begin{enumerate}[itemsep=0pt]
        \item We introduce three transition-aware locomotion adaptations that smooth locomotion initiation, turning, and stopping during VR navigation.
        \item We present a preliminary evaluation using objective balance measurements together with subjective assessments of balance confidence and workload.
        \item We provide initial evidence that locomotion adaptations applied during body-state transitions can improve postural stability without noticeably affecting perceived workload or simulator sickness.
    \end{enumerate}

\section{Related Work}
        Despite the advantages of immersive and interactive experiences in VR, virtual environments still face significant challenges related to motion sickness, balance, and spatial orientation during user interaction \cite{mahmud_2026_multimodal, mahmud_2024_multimodal, vo_2017_spatial}. Prior work evaluated the effects of prolonged VR exposure on motion sickness and postural instability, showing that extended VR use can increase oculomotor discomfort, disorientation, and postural instability \cite{pujiartati_2024_prosthetic_effects}. 
        
         Numerous approaches have been proposed to improve VR locomotion and user experience. Visual feedback techniques, including balance cues, wearable visual feedback, and multimodal feedback, have been shown to reduce center-of-pressure (COP) displacement and improve standing stability \cite{hossain_2026_cogniaudit, mahmud_2023_visualfeedback, mahmud_2022_auditory, mahmud_2022_standing, mahmud_2025_vibrotactile, Wang_2021_feedback}. Other work has focused on improving locomotion through rendering optimization, navigation assistance, and adaptive interaction. Examples include locomotion-aware foveated rendering \cite{shi_2023_foveated}, EEG-driven adaptive virtual environments \cite{chi_eeg}, and subtle visual or haptic navigation cues \cite{dubiel_subtle_cues}. These approaches demonstrate that adapting VR systems according to user state or interaction context can improve navigation performance, immersion, and overall usability.

        Despite extensive research on adaptive VR systems, existing approaches primarily focus on improving continuous locomotion, navigation efficiency, rendering performance, or reducing imbalance through visual feedback and environmental adaptation \cite{dubiel_subtle_cues, mahmud_2023_theeyes, mahmud_2024_multimodal}. While these methods successfully modify different aspects of the VR experience, they generally treat locomotion as a continuous process and apply adaptations throughout navigation rather than at specific moments of movement. However, VR navigation consists of a sequence of body-state transitions, and these transitions introduce abrupt changes in optic flow and movement dynamics that may require users to perform additional postural adjustments to maintain balance. Despite occurring repeatedly during navigation, these transition events have received comparatively little attention in the design of adaptive locomotion techniques. Our work addresses this gap by investigating whether adapting locomotion specifically during body-state transitions can improve postural stability while maintaining a comfortable navigation experience.

\section{Study Design}
    We evaluated the effectiveness of the proposed transition-aware locomotion adaptations by collecting objective and subjective measurements. We conducted a preliminary within-subjects pilot study with four healthy adults (2 male, 2 female, aged 26-33 years, M = 29.5). Participants had varying levels of prior VR experience, ranging from occasional to experienced users. All provided informed consent before participating in the study. 
    
    The study was conducted using an HTC Vive Focus Vision HMD in a standing VR navigation task. Participants stood on a BTrackS Balance Plate throughout the experiment and remained physically stationary while navigating through a virtual environment by locating three target locations using continuous hand-controller locomotion. Participants completed five locomotion conditions: No Adaptation (baseline), Motion Acceleration, Motion Deceleration, Turn Acceleration, and Transition Smoothing, which combined all three proposed adaptations. In the Motion Acceleration condition, forward locomotion speed increased linearly from 0\% to 100\% over 1 second when participants transitioned from the standing state to the walking state in the virtual environment rather than reaching the 100\% target forward speed instantaneously. In the Turn Acceleration condition, rotational speed increased linearly from 0\% to 100\% over 0.5 seconds when participants began turning, rather than reaching the 100\% target turning speed instantaneously. In the Motion Deceleration condition, forward motion decreased linearly from 100\% to 0\% over 0.5 seconds when participants transitioned from the walking state to the stopping state rather than stopping suddenly. If the user provided no locomotion input through the controller, the adaptation was not activated. The transition durations were chosen through pilot testing, as they produced smooth locomotion while maintaining responsive navigation. If the participant released the controller input before the transition was completed, the corresponding adaptation terminated immediately and locomotion returned to its normal state. The condition order was counterbalanced across participants to reduce learning and fatigue effects. Participants completed 6 trials per condition. Objective balance performance was assessed using the Balance Plate, which continuously recorded the participant's center of pressure (COP) during VR navigation at 25 Hz. From the recorded COP trajectories, mean COP velocity was computed for each trial and used as the primary objective measure of postural stability. Moreover, three subjective questionnaires were taken. Participants completed the Activities-specific Balance Confidence (ABC) Scale \cite{powell_1995_abc} and NASA Task Load Index (NASA-TLX) \cite{hancock_1988_nasa} after each locomotion condition. Simulator Sickness Questionnaire (SSQ) \cite{kennedy_1993_ssq} was taken before and after the experiment. 

    As the study was intended as a preliminary evaluation, the collected data were analyzed to identify trends in both objective and subjective measures. Mean COP velocity served as the primary evaluation metric, while the ABC Scale, NASA-TLX, and pre-post SSQ responses were used to assess perceived balance confidence, workload, and simulator sickness associated with the proposed transition-aware locomotion adaptations.

\section{Results}
    \subsubsection*{Objective Balance Performance}

\begin{figure}[H]
             \centering
             \includegraphics[width=\columnwidth]{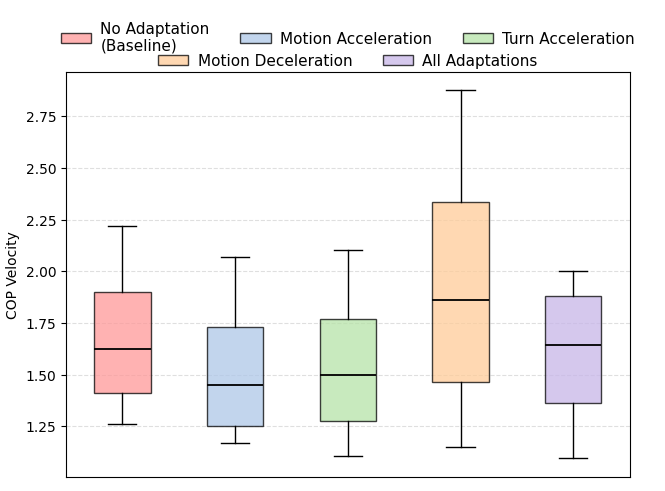}
             \caption{Distribution of COP Velocity Across Conditions}
             \label{fig:mean_cop}
        \end{figure}

        Figure \ref{fig:mean_cop} summarizes the mean COP velocity across the five locomotion conditions. Lower velocity indicates better postural stability. Motion Acceleration produced the lowest mean COP velocity among all evaluated conditions, corresponding to approximately 8.90\% lower COP velocity than the baseline condition. Turn Acceleration and the combined Transition Smoothing condition also reduced COP velocity relative to the baseline (7.95\% and 5.15\%, respectively), whereas Motion Deceleration  increased mean COP velocity by 15.10\%. Although the repeated-measures analysis did not reveal statistically significant differences across conditions (p $>$ 0.05), the observed trend consistently showed lower COP velocity for Motion Acceleration and Turn Acceleration, suggesting that gradually increasing locomotion speed and turning speed during movement and turning initiation may reduce postural instability during navigation.

        As shown in Table \ref{tab:completion_time_all}, all participants successfully completed the task under all conditions. Task completion times remained comparable across conditions. Although slight differences were observed between conditions, these variations were small relative to the inter-participant variability, suggesting that the proposed locomotion adaptations did not substantially affect navigation efficiency.
        \begin{table}[H]
            \centering
            \caption{Task Completion Time (in seconds) per Participant across Conditions.}
            \label{tab:completion_time_all}
            \small
            \resizebox{\columnwidth}{!}{%
            \begin{tabular}{lccccc}
            \hline
            \textbf{Participant} & \textbf{\shortstack{No \\ Adaptation}} & \textbf{\shortstack{Motion \\ Acceleration}} & \textbf{\shortstack{Turn \\ Acceleration}} & \textbf{\shortstack{Motion \\ Deceleration}} & \textbf{All} \\
            \hline
            P1 & 94.01  & 104.26 & 87.41  & 65.30  & 77.92  \\
            P2 & 202.72 & 108.70 & 156.50 & 150.70 & 133.22 \\
            P3 & 117.82 & 93.81  & 106.92 & 169.26 & 117.10 \\
            P4 & 75.27  & 191.52 & 198.79 & 90.91  & 169.76 \\
            \hline
            \textbf{Mean} & \textbf{122.45} & \textbf{124.57} & \textbf{137.40} & \textbf{119.04} & \textbf{125.00} \\
            \textbf{SD}   & \textbf{56.27}  & \textbf{45.06}  & \textbf{50.20}  & \textbf{49.00}  & \textbf{38.07}  \\
            \hline
            \end{tabular}%
            }
        \end{table}

    \subsubsection*{Subjective Evaluation}
\begin{figure}[H]
             \centering
             \includegraphics[width=\columnwidth]{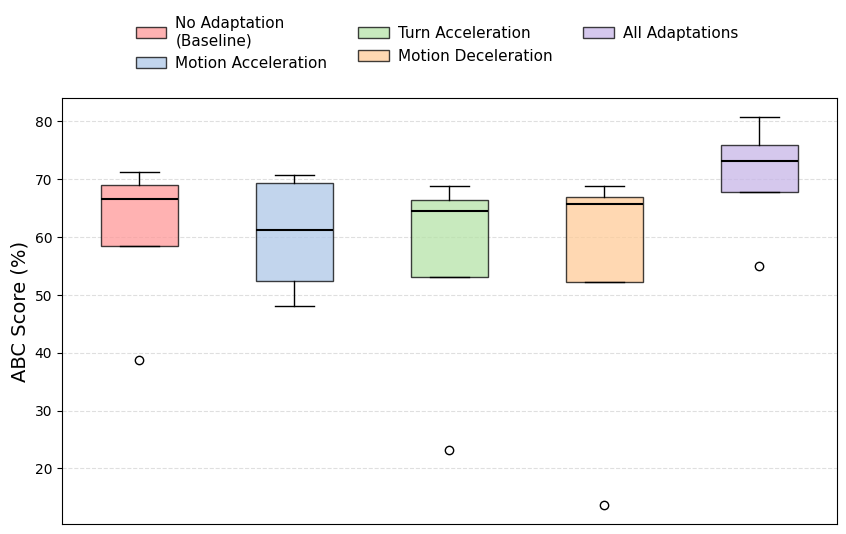}
             \caption{Distribution of Activities-Specific Balance Confidence (ABC) Scores across All Conditions}
             \label{fig:abc_scale}
        \end{figure}

        Participants reported greater confidence in maintaining balance while using the proposed transition-aware adaptations. Figure \ref{fig:abc_scale} presents the average ABC scores for each condition. The combined Transition Smoothing condition produced the highest average balance confidence among the adaptation conditions. Motion Acceleration also improved confidence relative to the baseline. The result is consistent with the objective balance measurements, indicating that the participants perceived greater stability when gradual locomotion transitions were introduced.

        NASA-TLX responses indicated that the proposed transition-aware adaptations did not consistently alter participants' perceived workload. While individual ratings varied across conditions, no systematic increase in mental demand, physical demand, effort, or frustration was observed. Moreover, the mean total SSQ score increased from $12.16 \pm 15.08$ before the experiment to $ 48.91 \pm 13.78$ after the experiment. Because SSQ was measured only before and after the complete session, its change reflects overall VR exposure. Combined with the objective reduction in COP velocity and the improved balance confidence reported in the ABC questionnaire, these findings suggest that the proposed adaptations can improve postural stability without imposing additional workload on users.

\section{Discussion and Conclusion}
    This preliminary study investigated whether adapting locomotion can support postural stability during body-state transitions. Although statistically significant differences were not observed because of the limited sample size, both the objective and subjective results indicate encouraging trends. Motion Acceleration and Turn Acceleration consistently reduced mean COP velocity compared with the baseline condition. Participants also reported higher balance confidence without noticeable increases in perceived workload. However, SSQ scores increased following the complete VR session. Because SSQ was not administered separately after each condition, this increase cannot be attributed to individual adaptations. One possible explanation is that locomotion initiation and turning represent periods of rapid sensory change. During these transitions, users experience sudden changes in optic flow while their physical bodies remain relatively stationary. This requires continuous integration of visual, vestibular, and proprioceptive information to maintain balance \cite{missen_2024_sensorysystem}. Previous research has shown that visually induced motion strongly influences postural control, and sudden visual perturbations can increase postural sway \cite{chander_2019_visualperturbations}. For this reason, our method adapts the locomotion behaviour only during transition periods rather than continuously throughout navigation. This suggests that selectively smoothing transitions may reduce unnecessary intervention while still supporting postural stability. 

    Interestingly, the three proposed adaptations did not produce equivalent improvements. Motion Acceleration achieved the largest average reduction in COP velocity (8.90\%), whereas Turn Acceleration demonstrated the largest within-subject effect size (Cohen's dz = 0.76), indicating a more consistent improvement across participants. In contrast, Motion Deceleration produced a negative effect size and increased mean COP velocity relative to the baseline. Some participants (P2, P3) reported that they felt slipping away in the environment because of Motion Deceleration. These observations suggest that the onset of locomotion and changes in walking direction may impose greater postural demands than locomotion termination. Rather than treating all movement transitions equally, future adaptive locomotion systems may focus on the transitions that affect balance the most.

    The primary limitation of this study is the small pilot sample of four participants, which limited the statistical power required to detect significant differences. To estimate the participant pool required for a future user study, we conducted an a priori power analysis using G*Power for a one-way repeated-measures ANOVA (within-subject design, five conditions) \cite{faul_Gpower_2007}. Assuming a medium effect size (Cohen's f = 0.25), significance level $\alpha$ = 0.05, statistical power = 0.80, one participant group, and five repeated measurements, the analysis indicated that 21 participants would be required for a sufficiently powered experiment. The observed pilot effect sizes ranged from 0.34 to 0.76, indicating small to moderately large effects. These findings support conducting a larger user study, although the current sample is insufficient for drawing definitive conclusions. Future work will recruit a larger participant population based on the power analysis and investigate the proposed adaptations across more diverse VR navigation tasks. In addition to COP velocity, future studies will incorporate eye-tracking and physiological measurements to better understand how body-state transitions influence visual attention, cognitive workload, and postural control. Overall, these preliminary findings suggest that adapting locomotion during body-state transitions is a promising direction to improve postural stability in VR navigation.

\balance
\bibliographystyle{abbrv}
\bibliography{template}
\end{document}